\def\epsffile#1{{\it Postscript Figure }#1}

\input epsf

\magnification= \magstep1                 
\tolerance=1600 
\parskip=5pt 
\baselineskip= 6 true mm \mathsurround=1pt

\font\medrm=cmr9

\def\secbreak{\vskip12pt plus 1in \penalty-200\vskip 0pt plus -1in} 

\def\Narrower{\par\narrower\noindent}   
\def\Endnarrower{\par\leftskip=0pt \rightskip=0pt} 
                
\def\a{\alpha}          \def\b{\beta}   \def\g{\gamma}  
            
            \def\k{\kappa}  \def\l{\lambda}          
                             \def\vv{\varphi}
\def\n{\nu}                 
\def\r{\varrho}         \def\s{\sigma}  \def\SS{\Sigma}

\def\w{\omega}  \def\W{\Omega}                          

 \def\LL{{\cal L}}

\def\cl{\centerline}    
\def\ni{\noindent}      \def\pa{\partial}       \def\dd{{\rm d}}        
\def\tl{\tilde}                        
\def\ra{\rightarrow}
\def\fn#1{\ifcase\noteno\def\fnchr{*}\or\def\fnchr{\dagger}\or\def
        \fnchr{\ddagger}\or\def\fnchr{\rm\S}\or\def\fnchr{\|}\or\def
        \fnchr{\rm\P}\fi\footnote{$^{\fnchr}$} {\scrunch#1 \toe} 
        \global\advance\noteno by 1 \ifnum\noteno>5\global\advance\noteno by-6\fi}
        \def\scrunch{\baselineskip=11 pt \medrm}
        \def\toe{\vphantom{$p_\big($}}
        \newcount\noteno

\def\fract#1#2{{\textstyle{#1\over#2}}}
\def\ffract#1#2{\raise .35 em\hbox{$\scriptstyle#1$}\kern-.25em/
        \kern-.2em\lower .22 em \hbox{$\scriptstyle#2$}}

\def\half{\fract12} 

\def\part#1#2{{\partial#1\over\partial#2}} 
 \def\ref#1{${\vphantom{)}}^#1$}
\def\ex#1{e^{\textstyle#1}}

\def\bbf#1{\setbox0=\hbox{$#1$} \kern-.025em\copy0\kern-\wd0
        \kern.05em\copy0\kern-\wd0 \kern-.025em\raise.0433em\box0}              

  \def\up#1{^{\vphantom{\Big]}\textstyle{#1}}}
  
\def\ref#1{${\,}^{#1}$}

\def\Gbar{\raise.13em\hbox{--}\kern-.35em G}
\def\lap{\setbox0=\hbox{$<$}\,\raise .25em\copy0\kern-\wd0\lower.25em\hbox{$\sim$}\,}
\def\glt{\setbox0=\hbox{$>$}\,\raise .25em\copy0\kern-\wd0\lower.25em\hbox{$<$}\,}
\def\gap{\setbox0=\hbox{$>$}\,\raise .25em\copy0\kern-\wd0\lower.25em\hbox{$\sim$}\,}

{\nopagenumbers %
\vglue 1truecm
\rightline{THU-97/14}
\rightline{gr-qc/9706058}
\vfil
\cl{{\bf THE SELF-SCREENING HAWKING ATMOSPHERE}\fn{Presented at
Strings97, Amsterdam, June 16-21, 1997.}\fn{This paper was modified
in order to attribute credit to Ref\ref6, where similar calculations
were reported.}} \medskip
\cl{A different approach to quantum black hole microstates}
\vfil

\cl{G. 't Hooft}
\bigskip
\cl{Institute for Theoretical Physics}
\cl{University of Utrecht, P.O.Box 80 006}
\cl{3508 TA Utrecht, the Netherlands}
\smallskip\cl{e-mail: g.thooft@fys.ruu.nl}
\vfil
\ni{\bf Abstract}\Narrower

A model is proposed in which the Hawking particles emitted by a
black hole are treated as an envelope of matter that obeys an
equation of state, and acts as a source in Einstein's equations.
This is a crude but interesting way to accommodate for the back
reaction.  For large black holes, the solution can be given
analytically, if the equation of state is $p=\kappa\rho$, with
$0<\kappa<1$.  The solution exhibits a singularity at the origin.
If we assume $N$ free particle types, we can use a Hartree-Fock
procedure to compute the contribution of one such field to the
entropy, and the result scales as expected as $1/N$.  A slight
mismatch is found that could be attributed to quantum corrections
to Einstein's equations, but can also be made to disappear when
$\k$ is set equal to one.  The case $\k=1$ is further analysed.

\Endnarrower
\vfil\eject}\pageno=1  
{\bf\ni 1. Introduction.}\medskip

Hawking radiation\ref1 is usually treated by means of a general
coordinate transformation between the frame of an infalling
observer and that of a stationary observer outside.  What is
obtained is the so-called Hartle-Hawking state, a state in which
the local energy-momentum tensor does not differ substantially
from that of the vacuum.  This Hartle-Hawking state, however,
does not yield any clues concerning the nature of the quantum
mechanical microstates of a black hole, even if arguments from
standard thermodynamics and statistical physics tell us what the
density of these microstates should be expected to be.

Recent results in string theory\ref2, complemented with the
higher-dimensional structures called $D$-branes\ref3, do show the
appearence of quantum microstates, but a detailed understanding
of these in terms of general properties of a horizon seems to be
missing.  What we intend to do is to establish to what extent one
can stretch more conventional versions of quantum gravity to
yield these quantum states.

The result of Ref\ref{4}, concerning the energy-momentum tensor
generated by Hawking radiation, will be interpreted in an unusual
fashion.  According to these authors, an anomaly in this tensor exactly
cancels out the contribution of Hawking radiation near the horizon.  In
this paper, we will assume that this anomaly is {\it not\/} present in
the energy-momentum tensor observed by the {\it outside\/} observer, but
only when this tensor is observed by observers in an infalling
(inertial) frame.  Further discussion of this issue is postponed to
Section~6.

For studying the complete set of quantum states, one must consider other
modes than just the Hartle-Hawking state.  The modes to be studied in
this paper are stationary in time.  In some sense, they are unphysical.
This is because a black hole in equilibrium with an external heat bath
cannot be stable; it has a negative heat capacity, and it is not
difficult to deduce that thermal oscillations therefore diverge.  Thus,
the states that will be discussed in this paper, being stationary in
time, are good candidates for the quantum microstates, but do not
represent the Hartle-Hawking state.  This way we claim to be able to
justify the calculations carried out in this paper:  we simply take the
energy-momentum tensor generated by Hawking radiation as if the Hawking
particles represented an `atmosphere'.  This atmosphere is taken as the
source for the gravitational field equations.  We ignore its
instability, which is relevant only for much larger time scales.
Thermal, non-interacting, massless particles obey the equation of state,
$$p=\fract13\r\,,\eqno(1.1)$$ 
where $p$ is the pressure and $\r$ the energy density.  If we have $N$
massless particle types, at temperature $T=1/\b$, we have
$$\r=3p={\pi^2\over 30}{N\over\b^4}\,.\eqno(1.2)$$ In particular, when
$N$ is large, the Hawking radiation is intense, and, at least far away
from the black hole, its effect on the space-time curvature should be
non-negligible.

In this paper, we regard Eq.~(1.1) or~(1.2) as describing matter near
the black hole, and at a later stage we will substitute the true Hawking
temperature for $\b^{-1}$.  Even if the reader finds there are reasons
to criticise these starting points, one may still be interested in
knowing the results of this model exercise.  Taking Hawking radiation as
a description of the boundary condition at some distance from the
horizon, we continue the solution of Einstein's equations combined with
the equation of state as far inwards as we can.  What is found is that
Hawking radiation produces radical departures from the pure
Schwarzschild metric near the horizon.  In fact, the horizon will be
removed entirely, but eventually a singularity is reached at the origin
($r=0$).  This should have been expected; our system is unphysical in
the sense that, when $M$ is sufficiently large, the Shandrasekhar limit
is violated, so that a solution that is regular everywhere, including
the origin, cannot exist.

In physical terms, what happens is the following (see Fig.~1).  A
blanket of Hawking particles near the (would-be) horizon is the source
of a gravitational field that is stronger that that of the entire black
hole, and therefore there will be gravitational field lines beyond the
blanket, pointing outwards, from the origin, rather than inwards.  On
the one hand, these field lines cause the gravitational potential to
rise when we follow it further inwards, so that the matter particles are
confined to stay close to the blanket, and on the other hand, in the
immediate vicinity of the origin, the field lines accumulate to form a
singularity there.  The singularity will be recognised as a
negative-mass singularity, and we will be able to compute precisely the
value of this negative mass (Eqs.~(3.13) and~(3.14)).

At this point it should be mentioned that after completion of an earlier
draft of this paper, the author learned that the differential equations
for this system, the so-called Tolman-Oppenheimer-Volkoff
equations\ref5, have been studied before by Zurek and Page\ref6, and
although they use slightly different variables, their conclusions were
essentially identical to the ones presented here.  In view of this, we
replaced the original Appendix (the contents of which can also be found
in Ref\ref6) by a more detailed elaboration of a special case.

\midinsert\cl{\epsffile{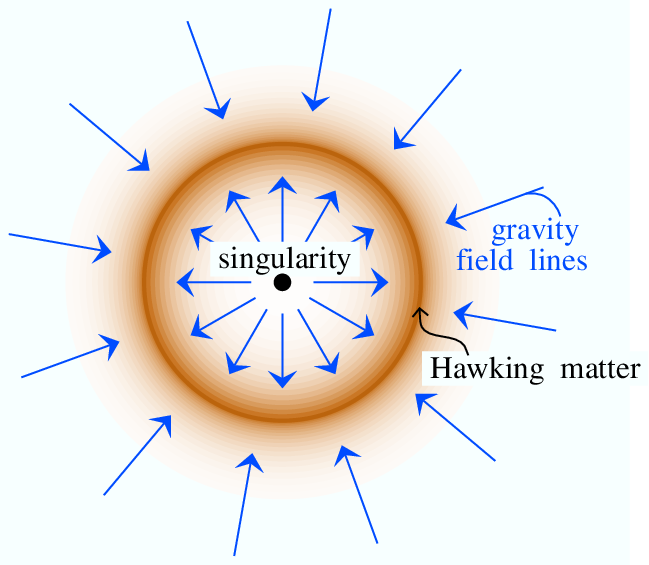}}
\cl{Figure 1. Sketch of configuration.}\endinsert
The negative-mass singularity is the only way in which this approach
departs (radically) from earlier Einstein-matter calculations\ref7.  But
for the remainder of our considerations this singularity is harmless.
It being repulsive, all particles will keep a safe distance from this
singularity.  This then, enables us to compute quantum effects in the
metric obtained.  Thus, in the second part of our paper, we compute the
entropy due to one of the scalar fields, using a WKB approximation.  We
find that this entropy is finite, so that the Hawking `blanket' itself
apparently acts as a soft alternative to the `brick wall' introduced in
Ref\ref{8}.  Furthermore, we find that the entropy per particle type
decreases as $N$ increases, simply because $N$ occurs in the source
described by Eq.~(1.2).  The total entropy of all particles becomes
independent of $N$, but is found to slightly overshoot Hawking's value.
We take this as a sign that the equation of state~(1.1) was too much of
a simplification.

It so happens that the metric can also be calculated if Eq.~(1.1) is
replaced by $$p=\k\r\,,\eqno(1.3)$$ where $\k$ is a coefficient ranging
anywhere between 0 and 1.  Moreover, the total entropy can also be
calculated in this case.  One may decide to adjust the value of $\k$
such that the entropy calculation matches precisely, but this could be
premature, because here one cannot ignore the quantum corrections to
Einstein's equations, since we are operating in the Plankian regime.  It
may nevertheless be of interest to note that a complete match is
achieved if $\k$ is set equal to 1, which is the case that will be
elaborated further in the Appendix.  It is a rather singular and
unphysical case.  We do conclude that, with interactions taken into
account, it may well be possible to obtain a self-consistent approach
towards microcanonical quantization of black holes, using ordinary
Hartree-Fock methods in standard gravity theories.  We do stress that
the price paid was a (mild) singularity at the origin.  A more thorough
analysis of the exact role played by this singularity in a more
comprehensive theory of quantum gravity is still to be performed.

\secbreak
{\bf\ni 2. The equations for $\k=\ffract13$}\medskip

We only consider spherically symmetric, stationary metrics in $3+1$
dimensions, of the form
$$\dd s^2=-A(r)\dd t^2+B(r)\dd r^2+r^2\dd\W^2\,,\eqno(2.1)$$
having as a material source a perfect fluid with pressure $p(r)$ and
energy density $\r(r)$. The Einstein equations read\fn{Here, 
units are chosen such that $G=1$.}
$$\eqalignno{1-\pa_r\left(r\over B\right)&=8\pi r^2\r\,,&(2.2)\cr
{\pa_r (AB)\over AB^2r}&=8\pi(\r+p)\,,&(2.3)}$$
where $\pa_r$ stands for $\pa/\pa r$.  The relativistic Euler
equations for a viscosity-free fluid can be seen here to amount to\ref{9}
$$\pa_rp=-(\r+p)\pa_r\log\sqrt{A(r)}\,,\eqno(2.4)$$
and the relation between $p$ and $\r$ is governed by an equation
of state. We will concentrate on the case
$$p=\k\r\,,\eqno(2.5)$$
where $\k$ is a fixed coefficient.  Massless, non-interacting particles
in thermal equilibrium have $\k=\ffract13$.  Although the calculations
described here can be performed for any choice of $\k$ between $0$ and
$1$, the special choice $\k=\ffract13$, appearing to be the most
important case, will be taken here first, because it shortens the
expressions considerably.  For the general case we refer to Ref\ref6.
If $\k=1$, however, complications arise, so this case is given special
attention in the Appendix.

Our liquid will be viscosity-free and free of vortices, 
so that Eq.~(2.4) can be integrated to yield 
$$8\pi\r A^2=C\,,\eqno(2.6)$$ 
where $C$ is a constant, later to be called $3\l^2/(2M)^2$.  After
inserting this equation into Eqs~(2.2)--(2.5), the latter can be cast in
a Lagrange form.  We will not discuss this Lagrangian however, but focus
on integrating the remainder of the equations.  What is needed there is
to observe the scaling behavior as a function of $r$.  Scale-independent
variables are $X$ and $Y$, defined by
$$\eqalignno{X& =A/r\,,&(2.7)\cr
\hbox{and}\qquad Y& =B\,.&(2.8)}$$
This turns the equations into:
$$\eqalignno{{r\pa_rX\over X}& ={CY\over 3X^2}+Y-2\,,&(2.9)\cr
{r\pa_rY\over Y}& =1-Y+{CY\over X^2}\,.&(2.10)}$$
and this allows us to eliminate $r$, obtaining a first order,
non-linear differential equation relating $X$ and $Y$.

The result of a numerical analysis of this equation is presented in Fig.~2.
Here, the constant $C$ has been scaled to one, by rescaling $X$. Each curve
now represents a solution.  For large $r$, all solutions spiral towards the
point $\W$.  For small $r$, only one curve allows $B$ to approach the value
one, so that, according to Eq.~(2.2), the density $\r$ stays finite, and
the metric~(2.1) stays locally flat.  This is the only regular solution.
Further away from the origin, this solution requires $\r$ always to be
large.

\midinsert\cl{\epsffile{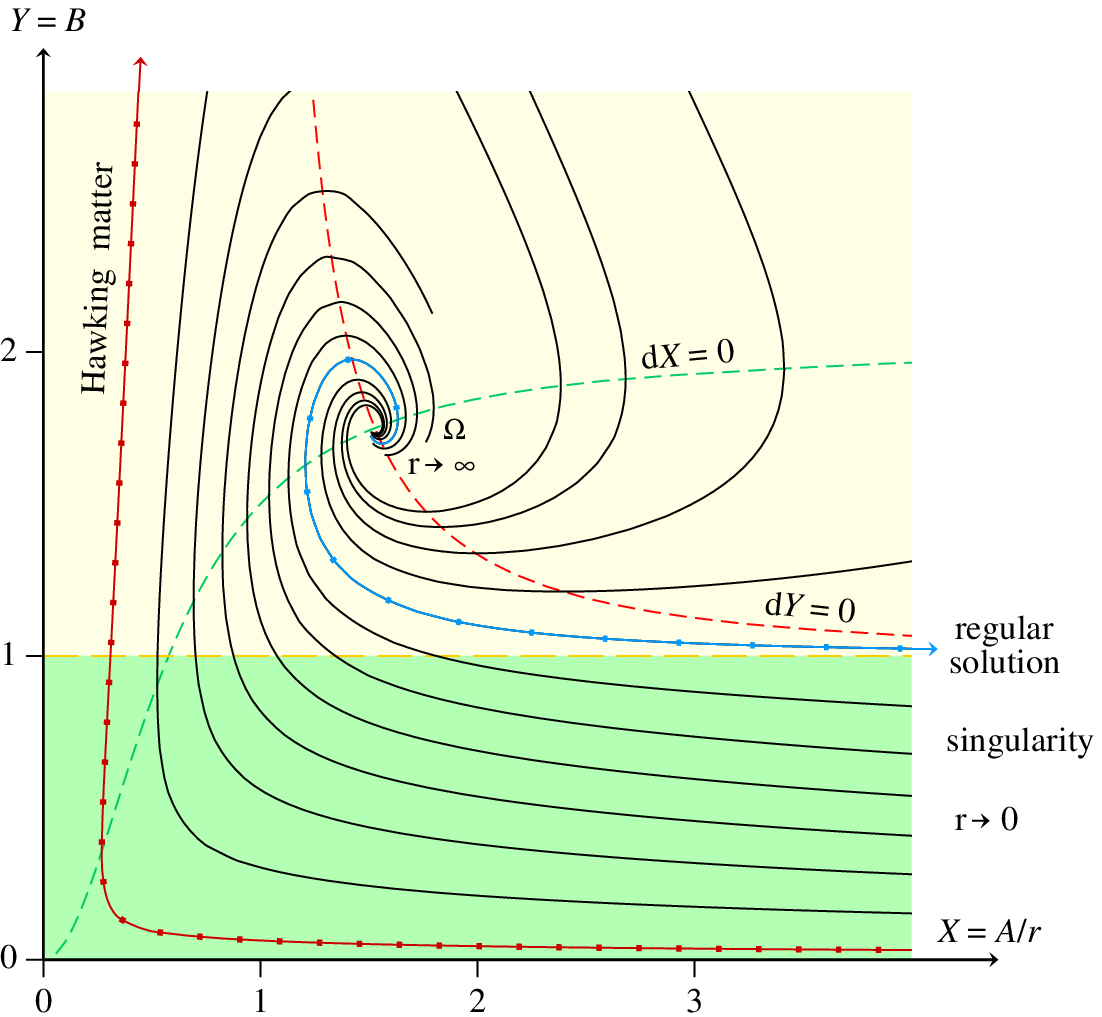}}
\cl{ Fig. 2. Solutions to Eqs. (2.9) -- (2.10). Solutions entering the}
\cl{shaded region must have a singularity at $r=0$.}\endinsert

We are interested in a very different class of solutions, the ones which,
far from the origin, approach a black hole surrounded by a very tenuous
cloud of matter, Hawking radiation.  This is the case where $Y\approx1$ and
$C\ll1$, or, after rescaling, $X\gg1$. We now observe that for very large
$r$, all solutions will eventually spiral into the point $\W$:
$$\W=\left(X=\sqrt{7C/3}\,,\ Y=\ffract74 \right)\,.\eqno(2.11)$$ 
Physically, this means that, since the universe is actually filled with
radiation, the curvature at large distances becomes substantial. For the
study of Hawking radiation, this large distance effect is immaterial and
will henceforth be ignored.

Thus, our boundary condition far from the origin will be chosen
to be
$$\eqalignno{A(r)&= 1-{2M\over r} \,,&(2.12)\cr
B(r)&=\left(1-{2M\over r}\right)^{-1}\,,&(2.13)\cr
8\pi\r(r)&={3\l^2\over(2MA)^2}\,,&(2.14)}$$
in the region $$\l\ll {r\over 2M}-1\ll {1\over\l}\,.\eqno(2.15)$$

Here, in Eq.~(2.14), the constant $C$ was replaced by $3\l^2/(2M^2)$,
since $\r$ has dimension $(\hbox{mass})\times(\hbox{length})^{-3}$
$=\,(\hbox{mass})^{-2}$, and the factor 3 is for later convenience.
In the mathematical construction of the equations, $\l$ will be
dimensionless.
If there are $N$ species of radiating fields, we have
$$ \l^2={4\pi^3\over45}{N\over\b^4}(2M)^2\,,\eqno(2.16)$$
where $\b$ is the inverse temperature. 
In Planck units, Hawking radiation has
$$\b=8\pi GM=8\pi M\,.\eqno(2.17)$$  
So, $$2M\l={1\over24}\sqrt{N\over5\pi}\,,\eqno(2.18)$$
and hence for large black holes, $\l$ is very small. 
The approximation~(2.15) holds over a huge domain.

When $r$ approaches the horizon, the effects of $\r$ are nevertheless
felt, and the solution becomes more complicated. Actally, $A$ never
goes to zero, so there is no horizon at all. Eventually, at small $r$,
$A$ diverges as a $\hbox{constant}/r$, which means that there is a
negative-mass singularity, as sketched in Figure~1.

Following the line of the solution of interest in Figure~2 (see also
Fig.~4), one gets the
impression that, for small enough $\l$, the solution may be found
analytically. This is indeed the case, as will be shown in the next
section.
\secbreak
{\bf\ni 3. The solution for small $\l$}\medskip
Eqs.~(2.9) and~(2.10) become slightly easier if we substitute $X$ and $Y$
by $P$ and $Q$:
$$\eqalignno{P&=(3/C)X^2/Y\,;&(3.1)\cr Q&=1/Y\,;&(3.2)\cr
\hbox{hence}\qquad X&=(C/3)^\half P^\half Q^{-\half}\,.\qquad\qquad{\ }
&(3.3)}$$
Defining $L=\log r$, the equations become
$$\eqalignno{\part PL&={3P\over Q}-5P-1\,;&(3.4)\cr
\part QL&=-{3Q\over P}-Q+1\,.&(3.5)}$$
In all limiting cases of interest to us, the r.h.s. of these equations will 
simplify sufficiently to make them exactly soluble.

Taking Eqs~(2.12) -- (2.14) to be valid in the region (2.15), for sufficiently
small $\l$, we can now integrate the equations down towards $r\ra0$. To do
this, we have to glue together different regions, where different approximations are
used. All in all, we cover the line $0<r\ll{2M\over\l}$ with six
overlapping regions, as depicted in Figure~3.
\midinsert\cl{\epsffile{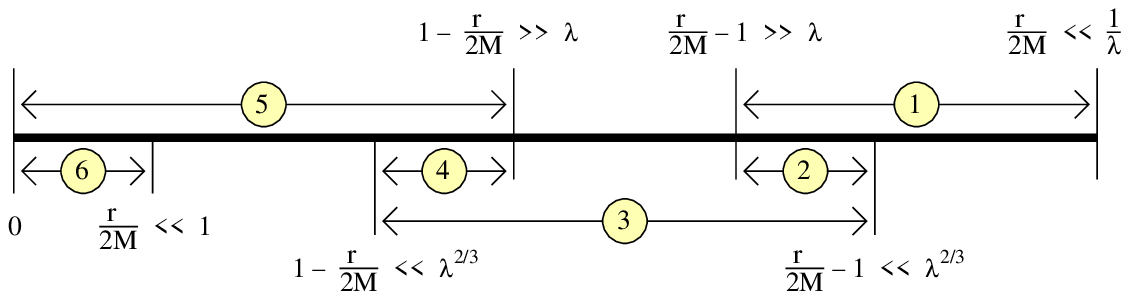}}
\cl{Fig. 3. Covering the line $0<r\ll 1/\l$ with regions (1) -- (6).}\endinsert

The region of Eq.~(2.15) is region~(1). The effects of $\r$ on the metric are
negligible, and so $A$ and $B$ still obey Eqs.~(2.12) and~(2.13). These
expressions agree with Eqs.~(3.4) and~(3.5) if we take
$P\gg Q$, so that (3.5) is readily integrated.

In region~(2) we have:
$$\displaylines{(2):\hfill\l\ll{r\over 2M}-1\ll\l^{\ffract23}\,.\hfill 
(3.6)}$$
It lies entirely inside region~(1), so here we have also:
$$r=2M+\s\ ;\quad A={\s\over 2M}\ ;\quad B={2M\over\s}\ ;
\quad 8\pi\r={3\l^2\over\s^2}\,.\eqno(3.7)$$
Region~2 is defined by the requirement $Q\ll P\ll 1$.

Region~3 has $P\ll 1$ and $Q\ll 1$. Here we can use $ {P\over Q}\equiv\w^2$
as a new variable. The differential relation between $L$ and $\w$ yields
$$\displaylines{(3):\hfill\eqalign{ r=2M\,\ex{\l\big(\w-{1/\w}\big)}\ ;&\qquad\quad
A=\l\w\ ;\qquad\ \cr 8\pi\r={3\over\w^2(2M)^2}\qquad;&\qquad\quad
B={\w^3\over\l(1+\w^2)^2} \,.}\hfill(3.8)}$$
The approximation corresponds to 
$$\left|{r\over 2M}-1\right|\ll \l^{\ffract23}\,.\eqno(3.9)$$
It is important to note that there are two integration constants in Eqs.~(3.8).
These have been adjusted such that agreement with region~(1) is obtained
where the two regions overlap, which is region~(2). 

Region~(4) is defined by $P\ll Q\ll 1$, and here the expressions (3.8)
simplify to 
$$\displaylines{(4):\hfill r=2M(1-P)\ ;\quad A={\l^2\over P}\ ;\quad
B={\l^2\over P^3}\ ; \qquad 8\pi\r={3P^2\over\l^2(2M)^2}\,.\hfill(3.10)}$$ 
This overlaps with region~(5), defined by $P\ll Q$ and/or
$Q\gg 1$).  In (5), integrating the equations yields
$$\displaylines{(5):\hfill\eqalign{{r\over 2M}= (1+5P)^{-\ffract15}\ \ ;&\qquad\quad
A={\l^2\over P} \left({2M\over r}\right)^6\,;\quad{\ }\cr B={\l^2\over
P^3}\left({2M\over r}\right)^{14}\ \ ;&\qquad\quad 8\pi\r={3P^2\over\l^2(2M)^2}
\left({r\over2M}\right)^{12}\,.}\hfill(3.11)}$$
Again the constants of integration have been carefully adjusted so as to
obtain agreement in region~(4), where (5) overlaps with (3).

Finally, in region~(6) we have $P\gg1$ and $Q\gg 1$, and our solution
simplifies into $$\displaylines{(6):\hfill\eqalign{
P={1\over5}\left({2M\over r}\right)^5\ ;&\qquad A=5\l^2\,{2M\over r}\ ; \cr
B=125\l^2\,{r\over2M}\ ;&\qquad 8\pi\r={3\over
25\l^2(2M)^2}\left({r\over2M}\right)^2\,.}\hfill(3.12)}$$

\midinsert \cl{\epsffile{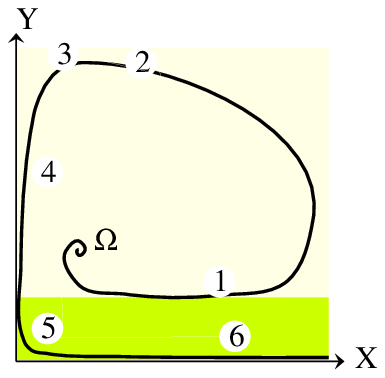}}\cl{Fig.~4. Sketch (not to scale) of the
path followed in the $X\ Y$ plane.}\endinsert
It is important that, throughout this calculation, the integration constants
in the different regions were carefully matched. So, now we can read off the
strength of the singularity at the origin. The space-like curvature
(determined by the $B$ coefficient) has the singularity of a ``black hole'' with
mass $$2M'=-{2M\over125\l^2}\,.\eqno(3.13)$$ The gravitational potential,
determined by $A$, is so much stretched in the time direction that it
corresponds to an object whose mass is only
$$2M''=-5\l^2\cdot 2M\,.\eqno(3.14)$$
Both masses are negative. \secbreak

{\bf\ni 4. The Rindler space limit}\medskip 

Let us now consider the limit where the black hole mass $M$ becomes
large. Assuming the surrounding matter to correspond to Hawking
radiation, we take $\l$ to obey Eq.~(2.18). From the 
inequalities in Figure~3 it may seem that, for $M$ large, $\l$ small,
region~(5) is a much larger stretch than region~(3), but, if the
geodesic distances are considered, the converse turns out to be true.
The distance between the point $r=2M$ (more or less in the middle of
region~(3)) and the origin at $r=0$ follows from Eqs.~(3.8):
$$\int_0^{2M}\sqrt{B(r)}\dd r=\int_0^12M\sqrt\l{\dd\w\over\sqrt\w}=4M\sqrt\l
=\sqrt{M\over 3}\left({N\over 5\pi}\right)^{\ffract14}\,,\eqno(4.1)$$ 
where, in the end, we inserted Eq.~(2.18). Notice, that the numbers are in
Plank units, so, if the black hole is large, the geodesic distance between
the apparent horizon $(r=2M)$ and our singularity at the origin is small in
comparison to the length scale of the black hole, but large in Planck units.

In contrast, region~(5) is small, in spite of the fact that $r$ runs
from $0$ to nearly $2M$. Let $\a$ be a number slightly bigger than $1$.
Then the geodesic length of the region between $r=0$ and $r=2M/\a$ is
$$\int_0^{2M/\a}\dd r\sqrt{B(r)}=2M\l\sqrt5\int_{\a^5}^\infty
{y^{\ffract15}\over(y-1)^{\ffract32}}\dd y\,,\eqno(4.2)$$
and this distance is of order of the Planck length, independently
of the mass $M$ of the black hole.

We now wish to consider the Rindler limit of our space-time, with which
we mean the limit $2M\ra\infty$, $2M\l$ fixed. In this limit, in
region~(3), space-time becomes flat (indeed, the matter density $\r$ in 
Eq.~(3.8) tends to zero). Only in region~(5) we get a deviation from 
the usual flat Rindler metric. Let us replace the coordinate $r$
by a more convenient coordinate $y$,
$$y=\left({2M\over r}\right)^5\ .\qquad\qquad(1\le y\le\infty)\eqno(4.3)$$
Furthermore, we replace the angular coordinate $\W$ by a transverse
coordinate $2M\tl x$. The metric, described by~(3.11), then becomes
$$\dd s^2\ =\ -{5\l^2 y^{\ffract65}\over y-1}\dd t^2\ +\ 5(2M\l)^2{y^{\ffract25}
\over(y-1)^3}\dd y^2\ +\ y^{-\ffract25}\dd\tl x^2\,.\eqno(4.4)$$
Substituting (2.18), and rescaling the time $t$, shows that this metric
will be universal for all black holes.

The gravitational potential $A(r)$ takes an absolute minimum inside
this region~(5):
$$\part Ar=0\qquad\ra\qquad  y=6\quad ,\qquad {r\over2M}={1\over\root
{\scriptstyle 5}\of 6}\,,\eqno(4.5)$$
and here, the matter density takes the extreme value
$$\r^{\rm extr}={3(y-1)^2\over200\pi(2M\l)^2}y^{-\ffract{12}5}=
{5\cdot 6^{\fract35}\over N}\,.\eqno(4.6)$$
Observe, that this is inversely proportional to the number of fields $N$
contributing to the Hawking radiation.
\secbreak

{\bf\ni 5. Free energy and entropy of a spectator field in this metric}
\medskip
In a metric of the form
$$\dd s^2=-A(y)\dd t^2+G(y)\dd y^2+H(y)\dd\tl x^2\,,\eqno(5.1)$$
a real scalar field with mass $m$ has a Lagrangian
$$\LL=\half H\sqrt{AG}\left(A^{-1}{\dot\vv}^2-G^{-1}
(\pa_y\vv)^2-H^{-1}(\tl\pa_x\vv)^2-m^2\vv^2\right)\,.\eqno(5.2)$$
Our calculation of the entropy goes as in Ref\ref{8}.
In the WKB approximation, at energy $E$ and transverse 
momentum $\tl k$, the oscillating part of the field $\vv(y)$ behaves as
$$\vv(y)\approx\ex{\ i\int k_y(y)\dd y}\,,\eqno(5.3)$$
with
$${k_y^2(y)\over G}={E^2\over A}-{\tl k^2\over H}-m^2\,.\eqno(5.4)$$
This expression transforms covariantly with general coordinate
transformations in the radial variable $y$, so the details of the radial
coordinate chosen are irrelevant.
We count the number $\n$ of solutions with energy below $E$, in a region
with transverse surface area $\SS$:
$$\pi\n=\sum_{\tl k}\int_Y k_y\dd y\,,\eqno(5.5)$$
where $Y$ is the domain of $y$ values in which the r.h.s. of
Eq.~(5.4) is positive. Writing $x=\tl k^2$, we find
$$\sum_{\tl k}=\SS\cdot\int{\dd^2\tl k\over(2\pi)^2}=
\SS\int_0^\infty{|\tl k|\dd|\tl k|
\over2\pi}={\SS\over4\pi} \int_0^\infty\dd x\,,\eqno(5.6)$$ and 
$$\n(E)={\SS\over(2\pi)^2}\int\dd x\sqrt{G\left({E^2\over A}-{x\over H}-
m^2\right)}\,.\eqno(5.7)$$
The integral over $x$ is over those values for which the argument 
of the root is positive,
$$\int_0^{P/Q}\dd x\sqrt{P-Qx}=\fract23{P^{\ffract32}\over Q}\,,\eqno(5.8)$$
so that
$$\n(E)={\SS\over 6\pi^2}{H\sqrt{G}\over A^{\ffract32}}
(E^2-m^2A)^{\ffract32}\,.\eqno(5.9)$$
The free energy $F$ at inverse temperature $\b$ is given by
$$\eqalign{\b F=&\sum_\n\log(1-e^{-\b E}) =\int_0^\infty\dd\n(E)\log(1-e^{-\b E})=
-\int_0^\infty\dd E\,{\b\n(E)\over e^{\b E}-1}\cr
=&-{\SS\b\over 6\pi^2}\int\dd y\int_{m\sqrt A}^\infty\dd E\,
{H\sqrt{G}(E^2-m^2A)^{\ffract32}\over
 A^{\ffract32}(e^{\b E}-1)}\,.}\eqno(5.10)$$
The entropy $S$ will be given by
$$S=\b^2\part F\b={\b\SS\over6\pi^2} \int\dd y\, {H\sqrt G\over A^{\ffract32} }
\int_{m\sqrt A}^\infty\dd E\,{(4E^2-m^2A)\sqrt{E^2-m^2A}\over e^{\b E}-1} 
\,.\eqno(5.11)$$

So-far, we left the mass of this field to be a free parameter, but from here
on, a considerable simplification is achieved by putting this mass equal to
zero (which is the case we are most interested in anyway). We can 
then proceed analytically. The integral over $E$ gives
$$m=0\quad\ra\quad S={2\pi^2\SS\over 45\b^3}\int\dd y\, {H\sqrt G\over
 A^{\ffract32}}\,.\eqno(5.12)$$

 In the Rindler limit, this integral will only receive a sizable contribution from
region~(5). Substituting the values for $A$, $G$ and $H$ that correspond to the metric
(4.4), we find that 
$$S={2\pi^2\SS\over45\b^3}{2M\over5\l^2}\int_1^\infty{\dd y\over y^2}\,,\eqno(5.13)$$
and with the Hawking temperature $\b=8\pi M$, this becomes
$$S={2\over 5N}\SS\,.\eqno(5.14)$$
If we add the contribution of all $N$ field types, the entropy per unite of
area becomes
$${S\over\SS}={2\over5}\,.\eqno(5.15)$$
This number should now be compared with what should have been expected,
the entropy of a black hole is
$${S\over\SS}=\ffract14\,,\eqno(5.16)$$
so we have a mismatch factor of
$${8\over5}\ .\eqno(5.17)$$

This number (which was also mentioned in Ref\ref6) is quite robust.  It
does not change if we change the number $N$ of the fields, or replace them
by vector fields.  It does change if either we add quantum corrections to
Einstein's equations, or if we modify the equation of state.  The latter is
an interesting exercise.  It turns out that the entire calculation given
above can easily be generalised into the case for general $\k$, as long as
$0<\k<1$.  Thus we reobtained the results of Ref\ref6.  The quantum modes
of the individual fields cannot be counted as easily as in the
non-interacting case, but it is not difficult to convince oneself that all
we really did was to integrate the entropy density of the material over the
curved space of our metric.  Now, the entropy density $s$ of matter at any
$\k$ value, is according to local observers,
$$s=\b(r)(1+\k)\r\,.\eqno(5.18)$$ Here, however, one has to substitute the
locally observed temperature, which is, due to redshift, given by
$$\b(r)=\sqrt{A(r)}\,\b\,,\eqno(5.19)$$ where $\b$ is the inverse
temperature as experienced by the distant observer.

The entropy for general $\k$ was also calculated by the author, but it 
also can be deduced from Ref\ref6, who use the coefficient $\g=\k+1$, 
or $n=1+{1\over\k}$. One finds for the total entropy $S$:
$${S\over\SS}={\k+1\over 7\k+1}\,,\eqno(5.20)$$
and this equals the desired value $\ffract14$ if $\k=1$.
It is difficult to imagine ordinary matter with such a high $\k$ value. 
Free massless fields generate the entropy density
$$ s=C\,N\,T^3\ ,\eqno(5.21)$$
where $N$ is the number of non-interacting field species. $C$ is a universal
constant. It appears that, at very high temperature, this should be replaced by
$$ s=C\,T^{1/\k}\ .\eqno(5.22)$$
A striking feature
of the result of Eq.~(5.20) is the independence on $N$. So, if $N$ is made
to depend strongly on temperature, this of course affects the equation of
state. If a strong kind of unification takes place at Planckian
temperatures, such that, at those temperatures $N$ suddenly decreases
strongly, one could imagine an effective increase of $\k$ beyond the
canonical value $\ffract13$. It is more likely, however, that this argument
is still far too naive, and that our approach must merely be seen as
a rough approximation. An error of 60\% is perhaps not so bad. On the other hand,
it is tempting to speculate that the case $\k\ra1$ has physical significance. 
This is a highly peculiar case, as is shown by the calculation in the (new)
Appendix. In this case, the total entropy receives its main contribution
from a very tiny region in space where the matter density reaches values 
diverging exponentially with the small parameter $\l$.\secbreak

{\bf\ni 6. Discussion and Conclusions}\medskip
Usually, it is argued\ref{6,10} that Hawking radiation produces an energy
momentum tensor that does not diverge at the horizon. The state of affairs
beyond the horizon, however, can be described in many ways, and depends
delicately on the nature of the quantum state considered\ref{11}. What we would
like to see, is an unambiguous prescription of the black hole quantum
states in terms of a general coordinate transformation of the observations
made by an inertial observer there.

The aim of this paper was to illustrate a point.  Suppose one assumed that
the majority of quantum states does show a strong back reaction of Hawking
particles upon the surrounding matter, just as if these particles form a
kind of ``atmosphere''\ref{12}. The price paid is a singularity, which in the
spherically symmetric case resides at the origin. But this singularity is
repulsive, and therefore its `quantum' effects are completely harmless.
One can calculate the entropy of this self-screening envelope of matter,
and the result is finite. Apparently, Hawking radiation can provide for its
own `brick wall'\ref{8}, in this case rather a `soft wall'.  The situation
is nearly stable; the entropy is larger than what it is supposed to be, but
only by a relatively insignificant numerical coefficient, $8/5$. It is
suggestive to speculate that this coefficient can be made to match, just by
adding non-trivial interactions. This would then provide us with a
highly non-trivial consistency check for the interactions up to the Planck
region, albeit that the check consists of just one numerical coefficient.

If the number of fields, $N$, is large, one may note that the distance
scale of the effects that lead to our entropy is large compared to the
Planck length (the length scale grows as $\sqrt N$), so that a fairly
consistent theory of black holes may result in the $N\ra\infty$ limit.

The role of the singularity should be further investigated. The picture
we get seems to tell us that particles with a large, negative masses may
have to be introduced in quantum gravity. A confinement mechanism
must assure that this negative mass is always surrounded by positive mass
material, so that we end up with a stable configuration. The rules for these
particles under general coordinate transformations should be further
investigated.

The question, `what does an ingoing observer see?', will be posed again and
again. Our (preliminary) answer is the following. Let us assume that, as
seen in the frame where the black hole was formed, and also in the frame of
the observer, indeed this observer survives his journey through the
horizon, until he or she hits the usual Schwarzschild singularity.  But
this is the black hole in the Hartle-Hawking state\ref{11}, and, as we
argued in the Introduction, this state fluctuates beyond control. Now,
since we put the black hole in a stationary, non-fluctuating state, we are
dealing with a superposition of many different Hartle-Hawking states, and
this superposition is determined by measurements made in the observer's
distant future.  Particles created by these measurements are likely to kill
the observer at the very moment he/she crosses the horizon. The complete
set of quantum states that serves as a basis in our configuration has a
vast majority of modes with energetic particles at the horizon that obstruct
a safe journey beyond.  Ordinary measurements are therefore not possible
there, and so it can happen that an entirely different description of this
region is called for. This description is the one presented in this paper.

\midinsert\cl{\epsffile{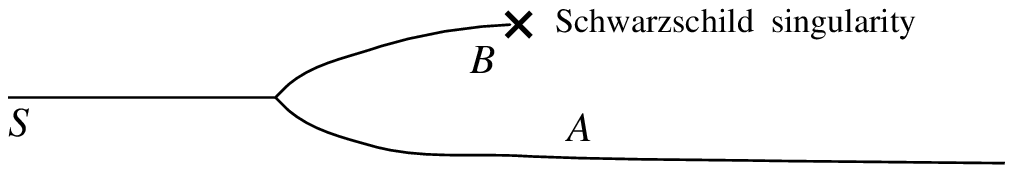}}
\cl{Fig. 5. The forked time parameter. Branch $A$ represents the outside
observer,}\cl{Branch $B$ the ingoing one.}\endinsert

An elegant way to formulate the situation may also be the following.
Since Hilbert space, as it is experienced by the infalling observer, is
constructed using states that are entirely different from the ones in
Hilbert space of the observer outside, it may be best to view the time
parameter $t$ in the Schr\"odinger equation as a parameter with multiple
``branches'', see Figure~5.  The Hamiltonian operator allows one to move
from one point to a neighboring point, but if one wishes to relate what
is seen by the outside observer, $A$, to the observations of the infalling
observer, $B$, one has to follow the Hamiltonian all around the forking
curve connecting the two observers.  Now, the energy-momentum operator is
directly related to energy, and the energy of a state is obtained by
Fourier transforming with respect to time.  Since the Fourier transform
of a time-dependent configuration along the curve connecting $S$ to $A$
may be different from the Fourier transform along the curve to $B$, the
expressions for the energy-momentum tensor for the two observers may be
different.  This holds for energy localized in some region (expressed by
the energy momentum tensor in that region); of course {\it total\/}
energy is the same everywhere since it is conserved.  This way one may
justify that the anomaly has to be taken into account for observer $B$
but not for observer $A$.

The case $\k=1$ is treated in detail in the Appendix. It is important 
to observe what happens in this case. The integral that yields the total entropy,
Eq.~(A.19), receives contributions that are fairly evenly distributed over
the range $0<r<2M$. But notice that the integrand is the product of an 
exponentially {\it large\/} factor, the entropy density (behaving as $\sqrt Q$), 
and an exponentially {\it small\/} factor, the space volume element, $\sqrt B$.
These functions both contain the factor $1/\l^2$ in the exponent 
(see Eqs.~(A.16)), and they both peak near the region
$r\approx0$. Thus, we have ventured deeply into the trans-Planckian region,
where the phytsical basis for these equations is extremely weak.
\secbreak

{\ni\bf APPENDIX.} {\it The equations for $0<\k<1$, and the calculation for 
the case $\k=1$}\medskip

For general $\k$, Eqs.~(2.6) and~(2.7) are replaced by
$$8\pi\r={\l^2\over \k(2M)^2}A\up{-{1+\k\over2\k}}\ ,\qquad\qquad
X=A\,r\up{-{4\k\over1+\k}}\,,\eqno(A.1)$$
and (2.9), (2.10) become
$$\eqalignno{{r\pa_r X\over X}&={\l^2 Y\over (2M)^2}X\up{-{1+\k\over2\k}}+Y-
{1+5\k\over1+\k} \,,&(A.2)\cr
{r\pa_r Y\over Y}&= 1-Y+{\l^2 Y\over\k(2M)^2}X\up{-{1+\k\over2\k}}\,.&(A.3)}
$$ It is convenient to define
$$\eqalignno{P&={(2M)^2\over\l^2}\,{X\up{1+\k\over2\k}\over Y}\,;&(A.4)\cr
Q&=1/Y\,.&(A.5)}$$
The field equations (3.4) and~(3.5) become
$$\eqalignno{{\dd P\over\dd L}&= {3\k+1\over2\k}\,{P\over Q}-{7\k+1\over2\k}P
-{1-\k\over2\k}&(A.6)\cr
{\dd Q\over\dd L}&=1-Q-{Q\over\k P}\,,&(A.7)\cr
\hbox{and}\qquad 8\pi\r&= {Q\over\k P r^2}\,.&(A.8)}$$

The solution of these equations can be obtained along the lines of
Sects.~3-5, but we now refer to Ref.\ref6. The identification relating
the variables $u$ and $v$ of Ref\ref6 with the $P$ and $Q$ used here is:
$$u={1-Q\over2}\ ,\qquad v={Q\over2P}\ ,\eqno(A.9)$$
and we find complete agreement with their Eq.~7, which corresponds to
our Eq.~(5.20). The black hole entropy is matched if one puts
$\k=1$. Let us now concentrate on this case.

When $\k=1$, Eqs.~(A.6) and (A.7) simplify into
$${\dd P\over\dd L}={2P\over Q}-4P\ ;\qquad{\dd Q\over\dd L}=1-Q-{Q\over 
P}\,.\eqno(A.10)$$
We have $$A=\l^2\left({r\over2M}\right)^2{P\over Q}\qquad\hbox{and}
\qquad B={1\over Q}\,.\eqno(A.11)$$
At sufficient distance from the black hole (but not too far away from it) we
have region~1, where one may assume that $P\gg Q$. 
There, we have Schwarzschild's solution:
$$P=\left({2M\over\l}\right)^2{1\over r^2}\left(1-{2M\over r}\right)^2\ ,
\qquad Q=A=\ 1-{2M\over r}\,.\eqno(A.12)$$
Near to the would-be horizon, region~1 overlaps with region~2, where
$Q\ll P\ll 1$, which in turn overlaps with region~3: $P\ll1$, $Q\ll 1$, and
there, one obtains $$Q={\l P\over\sqrt P-\l}\ ,\qquad L=\l\sqrt 
P+\l^2\log\left({\sqrt P-\l\over\l}\right)\approx 0\ ;\qquad A=\l(\sqrt 
P-\l)\,.\eqno(A.13)$$
Since in region~2 we have $P\gg\l^2$, we find that (A.12) and (A.13) agree 
in that region. Region~3 has  overlap with region~4, defined by $P\ll Q\ll1$, which
in turn overlaps with region~5, defined by
$P\ll1$ and $P\ll Q$. Here, it is convenient to use $Q$ as independent
variable, while solving for $P$ and $L$:
$$\eqalign{P&={\l^2 Q\over Q-\l^2(4Q\log Q+2)}\ ,\cr
L&=-\l^2\int{\dd Q\over Q-\l^2(4Q\log Q+2)}\ \approx 0\,.\cr}\eqno(A.14)$$
It overlaps with region~6 ($P\ll1$, $Q\gg1$), which in turn overlaps with
region~7, defined by $Q\gg1$ and $P\ll Q$. Taking $P$ as the indpendent variable,
one finds in region~7:
$$Q^4=P\ex{\,{1\over\l^2}-{1\over P}}\ ,\qquad L=-\fract14\log
\left({P\over\l^2}\right)\,,\eqno(A.15)$$ from which
$$\eqalign{P=\l^2\left({2M\over r}\right)^4\ ,&\qquad Q=\sqrt\l\left(
{2M\over r}\right)\ex{{1\over\l^2}\left[1-\big({r\over2M}\big)^4\right]}\,,\cr
A=\l^\fract72\left({2M\over r}\right)\ex{{1\over\l^2}\left[\big({r\over2M}
\big)^4-1\right]}\ ,&\qquad
B=\l^{-\half}\left({r\over2M}\right)\ex{{1\over\l^2}\left[
\big({r\over2M}\big)^4-1\right]}\,.\cr}\eqno(A.16)$$
Region~7 includes the singularity at the origin.
  
The entropy now cannot be computed directly by counting states, but we
can integrate the entropy density. From (5.18) and (5.19) we find that
the local entropy density is
$$s={\l^2\b\over8\pi\k(2M)^2}(1+\k)A\up{-{1\over2\k}}\,,\eqno(A.17)$$
so that in our case,
$${S\over\SS}=\l\int\left({r\over2M}\right)^2{\dd L\over\sqrt P}\,,\eqno(A.18)$$
and this yields
$${S\over\SS}=\int_0^{2M}\left({r\over2M}\right)^3{\dd r\over2M}=\ffract14\,.
\eqno(A.19)$$
 .
\secbreak
{\bf\ni References}\medskip
\item{1.}  S.W. Hawking, Commun. Math. Phys. {\bf 43} (1975) 199;     
W.G. Unruh, Phys. Rev. {\bf  D14} (1976) 870;      
W. Israel, Phys. Lett. {\bf 37A} (1976) 107.
\item{2.} J.M. Maldacena, {\it Black Holes in String Theory}, Princeton University
Dissertation, June 1996, hep-th/9607235, and references therein.
\item{3.} J. Dai, R. Leigh and J. Polchinski, Mod. Phys. Lett. {\bf A 4} (1989)
2073, J. Polchinski, Phys. Rev. Lett. {\bf 75} (1995) 4724, hep-th/9510017.

\item{4.} P.C.W. Davies, S.A. Fulling and W.G.~Unruh, Phys. Rev. {\bf D13}
(1976) 2720.

\item{5.} R.C. Tolman, {\it Relativity, Thermodynamics, and Cosmology\/} 
(Clarendon, Oxford, 1934), p.p. 242-243;
R.~Oppenheimer and G.~Volkoff, Phys. Rev. {\bf 55} (1939) 374.
\item{6.} W.H. Zurek and D.N. Page, Phys. Rev. {\bf D29} (1984) 628.

\item{7.} H. Bondi, Proc. R. Soc. London {\bf A282} (1964) 303; 
R.D. Sorkin, R.M.~Wald and Z.Z.~Jiu, Gen. Rel Grav. {\bf 13} (1981) 1127.
\item{8.} G. 't Hooft, Nucl. Phys. {\bf B256} (1985) 727  
\item{9.} L.D. Landau and E.M. Lifshitz, {\it Fluid Mechanics}, Pergamon Press, 
1959, p. 500.
\item{10.}  W.G. Unruh and R.M. Wald, Phys. Rev. {\bf  D29} (1984) 1047.
\item{11.}  J.B. Hartle and S.W. Hawking, Phys.Rev. {\bf D13} (1976) 2188.
\item{12.}  K.S. Thorne, {\it Black Holes: the Membrane Paradigm}, 
Yale Univ. Press, New Haven, 1986.

\bye